\documentclass[prl,aps,twocolumn]{revtex4}
\usepackage{amsmath}
\usepackage{amsfonts}
\usepackage{amssymb}
\usepackage{graphicx}
\usepackage{color}
\graphicspath{{images/}}

\def\beq{\begin{equation}}
\def\eeq{\end{equation}}
\def\bea{\begin{eqnarray}}
\def\eea{\end{eqnarray}}

\newcommand{\corr}[1]{\langle #1 \rangle}
\newcommand{\const}{{\rm const}\,}

\newcommand{\cT}{{\cal T}}

\begin{document}

\title{Excitation spectrum of a 2D long-range Bose-liquid with a supersymmetry}
\author{E. V. Mozgunov$^{1,2}$ and M. V. Feigel'man$^{1,2}$}
\affiliation{$^1$ L. D. Landau Institute for Theoretical Physics, Kosygin str.2, Moscow
119334, Russia}
\affiliation{$^2$ Moscow Institute of Physics and Technology, Moscow 141700, Russia}
\date{\today }

\begin{abstract}
We have studied excitation spectrum of the specfic 2D model of strongly interacting Bose particles
via mapping of the many-body Schrodinger equation in imaginary time  to the classical stochastic dynamics.
In a broad range of coupling strength $\alpha$ a roton-like spectrum is found, with the roton gap being extremely
small in natural units. A single quantum phase transition between strongly correlated supefluid and quantum
Berezinsky crystal is found.
\end{abstract}

\maketitle

Usually the system of Bose particles at zero temperature exists in one of two possible ground-states:  superfluid  (SF) or crystalline (CR). More exotic option is a "supersolid" ground state suggested long ago~\cite{AndreevLifshitz},  which
attracted a lot of attention recently~\cite{supersolid-new}; this is a state which is expected to possess \textit{both superfluid and
crystalline} order simultaneously. Another direction of the search for unusual quantum ground states is related
with a search for a "Bose-metal", that is, a bosonic analog of a Fermi-liquid, see for example \cite{DasDoniach,Paramekanti}.
 Such a state would possess \textit{neither superfluid no crystalline} order.
Suggestion for the search of such a strange quantum state was made 20 years ago in Ref.~\cite{fgv}, in relation with classical thermodynamics of 3D vortex liquid in high-temperature superconductors.
This idea was further developed in Ref.~\cite{fgil} where two different models of strongly interacting Bose-liquid were considered (note that Refs.~\cite{fgv,fgil} refer to continuous 2D Bose-liquids without any lattice, whereas Refs.~\cite{DasDoniach,Paramekanti} consider lattice models).
The arguments were given in Ref.~\cite{fgil} in favor of existence of a new unusual
ground-state which is still {\it liquid}, but is {\it not superfluid}.  One of these models refers to 2D bosons interacting with a 2D dynamic $U(1)$ gauge field, with an effective coupling constant $\sim 1$. The second model (KKLZ, Ref.~\cite{kklz}) is purely static, it has a remarkable feature that its exact ground-state wavefunction is represented in a simple Jastrow form.

It was shown later in Ref.~\cite{FS} that KKLZ model obeys nonrelativistic supersymmetry which allows to obtain a number of interesting results analytically.
The KKLZ model contains a coupling
constant $\alpha$ such that small values $\alpha \leq 3$ definitely lead to a gapful superfluid state, whereas at very large
$\alpha \geq 35$ a kind of a "Berezinsky crystal" with power-law decay of positional correlations is stabilized, according
to Ref.~\cite{wigner}.
An issue was raised in Ref.~\cite{FS} about possible existence of a third\underline{,} intermediate\underline{,} ground-state of the "normal liquid" type, which could exist in some part of the broad range $3 \leq \alpha \leq 35$.
Supersymmetry of KKLZ model makes it also possible to compute time-dependent quantum correlation  functions via classical Langevin dynamics (the relation between supersymmetry and Langevin dynamics was discussed, in particular,
in Ref.~\cite{ft}).
 Similar approach was proposed by C.Henley~\cite{Henley2004} for the lattice
quantum systems and used efficiently in Ref.~\cite{Qdimers} to explore excitation spectrum of quantum dimer models on square and triangular lattices at the Rokhsar-Kivelson point~\cite{RoksarKivelson}. More recently, the same lines of ideas were developed in Ref.~\cite{Castelnovo} for quantum spin models.

In the present Letter we report results of extensive numerical studies of dynamic density-density correlation function
 in the  KKLZ model through a broad range of coupling strength $2 < \alpha < 40$.  The presence of roton-like branch
 of the excitation spectrum is demonstrated, with the ratio of the roton gap $\Delta$ to the plasma frequency $\omega_0$
strongly decreasing with increase of $\alpha$. Right before the crystallization transition at $\alpha =\alpha_c \approx 37$, this ratio  becomes less
than $ 3\cdot 10^{-3}$; still we could not identify any finite interval of $\alpha < \alpha_c$
where roton gap would be exactly zero without a Berezinsky crystal being formed.
An effective roton mass $m^*$
defined via the spectrum $\omega(p) = \Delta + (p-p_0)^2/2m^*$ near the roton minimum, is found to be weakly-dependent upon $\alpha$.  The
spectral weight $S(p,\omega)$ is well-approximated by the single quasiparticle peak at $\omega < 2\Delta$, whereas at higher energies
quasiparticle spectrum is undefined due to strongly decaying nature of excitations. Our results support an existence of superfluid ground-state all the way up  to the crystallization transition, but the transition temperature $T_c(\alpha)$ scales with $\Delta(\alpha)$
and becomes extremely low at $\alpha$ close to $\alpha_c$.

We study the KKLZ model of 2D interacting Bose-particles characterized by the exact
ground-state wavefunction of Jastrow form
\beq
  \Psi_0({\bf r}_1,...,{\bf r}_N) = \const \cdot
  \prod_{j>k}|{\bf r}_j-{\bf r}_k|^{2\alpha}
  e^{-\pi\alpha n \sum_i r_i^2}.
\label{psi} \eeq Here $n$ is the particle density and $\alpha$ is a parameter. Many-body probability density $P_0({\bf r}_1,...,{\bf r}_N) =
|\Psi_0({\bf r}_1,...,{\bf r}_N)|^2$ can be considered as a Gibbs measure for a classical 2D liquid with potential energy \beq \label{V}
  V\{{\bf r}_i\} =
    - 4\alpha\cT \sum_{j>k} \ln|{\bf r}_j-{\bf r}_k|
    + 2\pi\alpha\cT n \sum_i r_i^2.
\eeq and temperature $\cT$.  Quantum Hamiltonian of the KKLZ model is defined as
\beq \hat{H} =  \sum\limits_j\sum\limits_{\mu=x,y}\hat{q}^{\dag}_{j,\mu} \hat{q}_{j,\mu} \, ; \quad \quad \hat{q}_{i,\mu} \equiv
i\hbar\frac{\partial }{\partial r_{j,\mu}} + \frac{i}{2}\frac{\partial V}{\partial r_{j,\mu}} \label{H1} \eeq
where we put $\cT = \hbar$ and $2m=1$.
 Langevin dynamics leading to
the Gibbs distribution $ P_0({\bf r}_1,...,{\bf r}_N)$ is defined as \beq \label{Langevin}
 \frac{dr_{j,\mu}}{dt} =
    - \frac{\partial V\{{\bf r}_i\}}{\partial r_{j,\mu}}
    + \xi_{j,\mu}(t)
\eeq
where $\overline{ \xi_{j,\mu}(t) \xi_{k,\nu}(t') } =
    2 \cT \delta_{jk} \delta_{\mu\nu} \delta(t-t')$.
Our goal is to compute  dynamic density-density correlation function $S(\mathbf{k},t) = \frac1{\cal V}\corr{n_\mathbf{k}(t)n_{-\mathbf{k}}(0)}$,  ($\cal V$ is the system's volume)  in the ground  state (GS) of  the Hamiltonian (\ref{H1}).
In terms of spectral expansion it is given  by
$S(\mathbf{k},t) = \sum_i \left| \langle GS|n_\mathbf{k}|\mathbf{k},i \rangle \right|^2 e^{-i\omega_{\mathbf{k},i} t}$ where $i$ denotes all
quantum numbers except the momentum $\mathbf{k}$. The equivalence~\cite{ft,FS,Henley2004,Castelnovo}
of quantum and classical dynamics for the
theories like the one defined by Eq.(\ref{H1}) allows us to use classical simulation of the Langevin dynamics defined in Eq.(\ref{Langevin}) to
compute $S(\mathbf{k},t)$ in the imaginary-time domain: $S(\mathbf{k},-i\tau) = \mathcal{S}(\mathbf{k},\tau)$, where
 \bea \mathcal{S}(\mathbf{k},\tau) = \int \prod_jd{\bf r}_j (0)d{\bf r}_j (\tau)\times \\
\times P({\bf r}_j(0),0,{\bf r}_j(\tau),\tau) \frac{1}{\cal V}\sum\limits_{i,j} e^{ikr_i(\tau)} e^{-ikr_j(0)}
\label{Equiv}
\eea
 where $ P({\bf r}_j(0),0, {\bf r}_j(\tau),\tau) $ is the 2-time $N$-particle joint distribution function for the
stochastic diffusion  process defined by Eq.(\ref{Langevin}). For the derivation of Eq.(\ref{Equiv}) see the Supplementary
material.

\begin{figure}[h]
\includegraphics*[width=8cm]{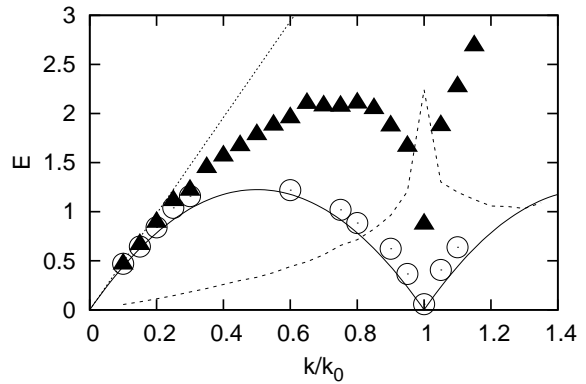}
\caption{Computed lower bound of the spectrum $E_{\rm min}(k)$ for $\lambda=2$ CSM is shown by circles, the results for simple spectral average $\omega_{fm}(k)$ are
shown as triangles, exact theoretical lower bound $E_-(k)$ is represented by the full line, dashed line shows static structure factor $S(k)$.} \label{Fig:CSM}
\end{figure}

We begin with the application of our computational method to the simpler case of the Calogero-Sutherland model (CSM) \cite{CSM} defined on a 1D
circle of the length $L$. The CSM ground-state wavefunction is $\psi_0=\prod_{i<j} \sin^{\lambda}({\pi x_{ij}/L})$, where $\lambda > 1/2$ and
arbitrary otherwise. The corresponding classical potential energy is $V_{CSM} = -\lambda \sum_{i<j} \ln\sin^2{\pi x_{ij}/L}$. We simulate CSM
model with $N= 200$ particles via Langevin dynamics to compute its dynamic structure factor ${\cal S}(k,\tau)$ and
 compare with exact results
available~\cite{Pustilnik}. According to Ref.~\cite{Pustilnik}, CSM spectral density $S(k,\omega) = \sum_i \left| \langle GS|A_k|k,i \rangle
\right|^2 \delta(\omega_i-\omega)$ is nonzero in a finite region $\omega\in[E_{-}(k),E_{+}(k)]$ only, where $E_{-}=v_s \left(k -
\frac{k^2}{k_0}\right),~ E_{+}=v_s \left(k + \frac{k^2}{\lambda k_0}\right) $, and $v_s = \pi \lambda \frac{n}{m}$ and $k_0=2\pi n$. In
Fig~\ref{Fig:CSM} we plot results of numerical simulation for $\lambda=2$ together with theoretical low bound curve. In our
computation, the lower bound of the spectrum was determined as the extrapolation $E_{\rm min}(k) = \lim_{t \to \infty} d\ln \mathcal{S}(k,t)/dt$; another
spectral characteristic is its simple average $\omega_{fm} (k) = \int_0^{\infty}\omega S(\omega,k) d\omega =
d\ln \mathcal{S}(k,t)/dt\vert_{t \to 0}$.
The agreement between data for $E_{\rm min}(k)$ and theoretical spectral boundary $E_-(K)$ is remarkable.
It proves the capability of our method to capture gapless excitations with large wavevectors $k \sim k_0$,
 which are invisible  in the "first moment" approximation $\omega_{fm}$. Note that for small $k \ll k_0$ data
for $E_{\rm min}(k)$ and $\omega_{fm}(k)$ coinside, as it should be for the spectral density nearly saturated by
single-particle excitations.

\begin{figure}[h]
\includegraphics*[width=8cm]{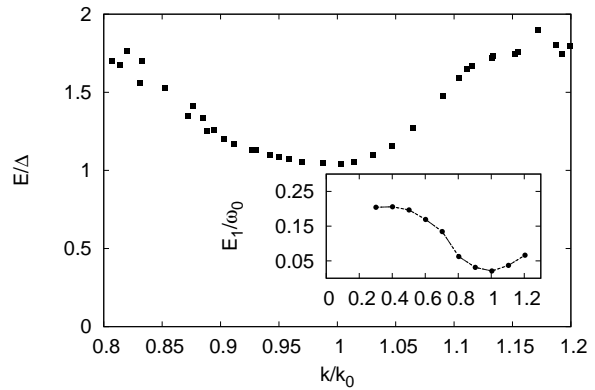}
\caption{Quasiparticle energy near the roton minimum for $\alpha =20$ obtained by
the fit of $\mathcal{S}(k,\tau)$; in the inset lower bound for $E(k)$ in the whole $k$-range is shown.}
\label{Fig:a20}
\end{figure}

Now we turn to our major subject: search for the low-energy roton modes in the KKLZ model defined by the Hamiltonian (\ref{H1}).
An example of the excitation spectrum in the strong coupling region,
$\alpha=20$, is shown in Fig. ~\ref{Fig:a20}, here and below $N=256$.
 We plot here the data for $E_{\rm min}(k)$ for the wavevectors
$k$ in the vicinity of $k_0 = 2\pi \sqrt{n}$, where static structure factor $S(k,t=0)$ has a peak.
The inset to Fig~\ref{Fig:a20} shows $E_{\rm min}(k)$ in a broad range of $k$ determined via best fit of
 $\mathcal{S}(k,\tau)$ to the single exponent $Ae^{-E_1\tau}$.
In the main panel of  Fig. ~\ref{Fig:a20} we show $E_{\rm min}(k)$ in the narrow region around $k_0$, obtained
via more accurate fiting procedure described in Suppl.2.

A roton minimum in $E_{\rm min}(k)$ is clearly visible at $k=k_0$; below we denote the roton gap as $\Delta =E_{\rm min}(k_0)$. For $\alpha=20$
the magnitude of the roton gap $\Delta$ is found to be very small, about $~1\%$ in comparison with the plasma frequency $\omega_0 = 4\pi\alpha
\frac{n}{m}$, which sets a natural energy scale in the problem.  In particular, $\omega_0$ is the frequency of the uniform density oscillations
in the KKLZ model, see Ref.~\cite{FS} for details. Thus, our first qualitative observation is that in the strong-coupling region the excitation
spectrum shows a very deep roton minimum. As follows from the general arguments~\cite{Pitaevsky}, a well-defined excitation spectrum may not
exist in the $k$ region where  quasiparticle decay is allowed by conservation laws. For the roton-like spectrum with deep minimum, the
"no-decay" condition is fulfilled at energies $ E < 2\Delta$ only:  at higher excitation energy, the  decay into two rotons is allowed with a high rate. A well-defined roton excitations may
exist in the momentum range $p_- < k < p_+$ around the minimal point $k_0$. According to Ref.~\cite{Pitaevsky}, the excitation energy $E(k)$ is
expected to approach the end-points $p_\pm$ nonanalitically, with a zero slope:
\beq E(k \to p_\pm) = 2\Delta - a e^{-b/|k-p_{\pm}|}
\label{stp}
\eeq
 where $p_{\pm}$ are called spectrum terminating points, and $a$ and $ b$ are some positive constants.
The equation (\ref{stp}) results~\cite{Pitaevsky}  from an exact summation of the most singular
diagrams for the momenta $k \approx p_\pm$.
 Our
data presented in Fig.~\ref{Fig:a20} (main panel) are in good qualitative agreement with this prediction; the spectrum end-points are situated
at $(p_-,p_+) \approx (0.8 , 1.2 )\cdot k_0 $. Unfortunately, high-presision computation of $E_{\rm min}(k)$ close to the end-points was found to the
very difficult due to increasing data scattering.

\begin{figure}[h]
\includegraphics*[width=8cm]{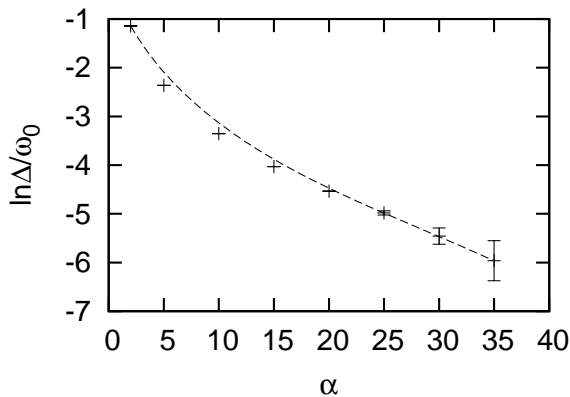}
\caption{Roton gap $\Delta$ as function of $\alpha$, logarithmic scale}
\label{Fig:gap1}
\end{figure}

\begin{figure}[h]
\includegraphics*[width=8cm]{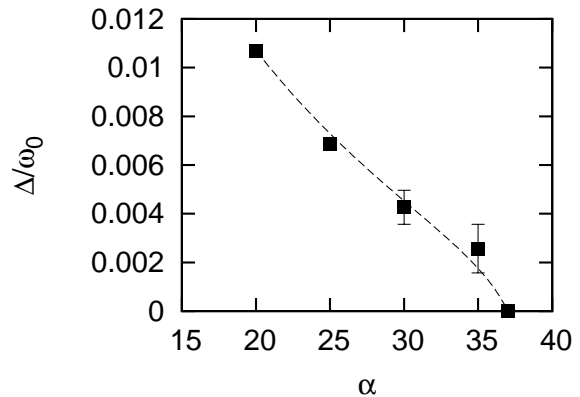}
\caption{Ratio of the roton gap to plasma frequency in the large-$\alpha$ range.}
 \label{Fig:gap2}
\end{figure}

Similar analysis of the relaxation data for different vaues of the coupling constant $\alpha$ yields the
dependence of the gap magnitude $\Delta$ on $\alpha$ presented in Fig.~\ref{Fig:gap1} in logarithmic scale.
Increase of $\alpha$ leads to very sharp (nearly exponential in the range $10 < \alpha < 35$) decrease of the  gap
 magnitude $\Delta$.  The same data for the region of large $\alpha \geq 20$ are presented in Fig.~\ref{Fig:gap2}
in linear scale. These results are consistent with linear vanishing of the gap at $\alpha \approx 37\div 38$,
slightly above the point of the crystallization transition $\alpha_{old} = 35$
found in Ref.~\cite{wigner} for classical 2D Coulomb gas.
However, the values of
$\Delta$ in this range contain large relative errors which makes it difficult to determine unambigously  where  $\Delta(\alpha)$ vanishes. To
approach the problem of location of the quantum critical point from another perspective, below we compare long-time asymptotics of the dynamic
structure factor $\mathcal{S}(k_0,\tau)$ in the liquid and crystalline phases.
\begin{figure}[h]
\includegraphics*[width=8cm]{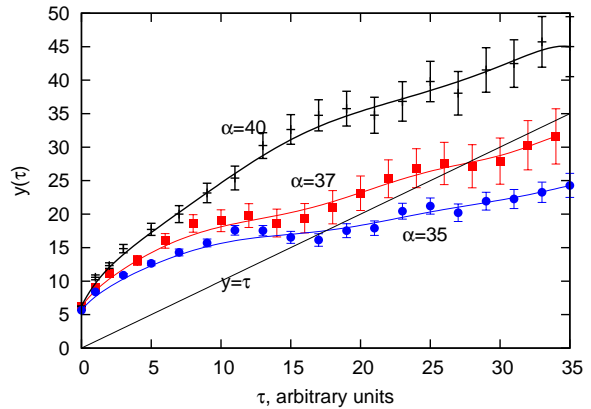}
\caption{Inverse logarithmic derivative of
$\mathcal{S}(k_0,\tau)$ is plotted for $\alpha$ = 35 (blue dots), 37 (red boxes) and 40 (black bars)
as measured directly in the simulation.}
\label{Fig:cri}
\end{figure}
The crystalline phase of the KKLZ model is very specific. This is densly packed triangular lattice, but, instead of usual transverse phonons
with $\omega \sim q$, it supports phonons with parabolic dispersion, $\omega(q) \sim q^2$. This comes from the fact that shear modulus of this
lattice vanishes itself in the long-wavelength limit, $\mu(q) \propto q^2$, see Ref.~\cite{FS}; here wavevector $\mathbf{q} = \mathbf{p} -
\mathbf{G}_i$, where $\mathbf{G}_i$ is one of principal inverse lattice vectors. The presense of soft shear modes leads to a specific long tail
in the time decay of the angle-averaged
 structure factor $\mathcal{S}(k_0,\tau) = \int \frac{d\varphi}{2\pi} \mathcal{S}(k_0\cos\varphi,k_0\sin\varphi,\tau)$.
which can be measured by Langevin dynamics:
\beq
-\frac{d\hbox{ln}\mathcal{S}(k_0,\tau)}{d\tau}=\frac{1}{2\tau}- f(\tau), ~ ~\tau >>\frac{m^*}{k_0^2}
\label{Scr}
\eeq
where $f(\tau)>0$ decays exponentially with $\tau$ and $m^*$ is the effective mass (to be discussed later).
Now we define a function  $y(\tau)=-\left(2 d\hbox{ln}\mathcal{S}(k_0,\tau)/d\tau\right)^{-1}$ and note that
according to Eq.(\ref{Scr}) it should never cross the line $y = \tau$. On  the other hand,
 in the liquid phase with a nonzero gap $\Delta$, the function $y(\tau)$
approaches $1/2\Delta$ at $\tau \to \infty$, so its crossing with the straight line $y = \tau$ occurs definitely.
In Fig~\ref{Fig:cri} we present simulation results for the function $y(\tau)$  at  $\alpha=35, 37$ and $ 40$.
According to the criterion formulated above, the critical value $\alpha_c$ is also found  in the range $37 <\alpha_c <38$.
 The data summarised in Fig.~\ref{Fig:gap2} and Fig.~\ref{Fig:cri}
support the conclusion that  liquid state with a small roton gap $\Delta$ transforms into a crystalline state
via the single phase transition where $\Delta$ vanishes.

\begin{figure}[h]
\includegraphics*[width=8cm]{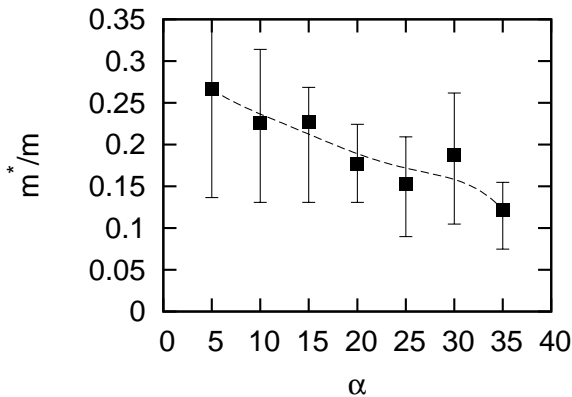}
\caption{Quasiparticle mass $m^*$  weakly depends on $\alpha$ even at the transition point.}
 \label{Fig:mms}
\end{figure}

Coming back to the discussion of the the gapful liquid phase at $\alpha<\alpha_c$, we note that
low-lying excitation with $k \approx k_0$ are chatacterized, apart from the gap value $ \Delta$, by the
value of the effective mass $m^* = \left(d^2E(k)/dk^2|_{k_0}\right)^{-1}$. Measurement of the
$S(k,\tau)$ decay in the vicinity of $k_0$ allows to determine $m^*$ in a broad range of $\alpha$,
as shown in Fig.\ref{Fig:mms}.
The results shown in Fig.~\ref{Fig:gap2} and Fig.~\ref{Fig:mms} yield the parameters of the
low-lying excitation spectrum $\varepsilon(k) = \Delta + \frac{(k-k_0)^2}{2m^*}$. allowing
to determine the temperature of superfluid-to-normal transition $T_c(\alpha)$.
Within the Landau-type  mean-field theory  $T_c$ is defined as the temperature where
superfluid density $n_s = n - n_n$  vanishes. Neglecting quasiparticles interaction,
we find equation for the critical temperature $T_c$:
 \beq
 n  = -\frac{1}{2m}\int\frac{\partial f_B(\varepsilon,T_c)}{\partial \varepsilon} k^2\frac{d^2 k}{(2\pi)^2},
\label{nn}
 \eeq
where $f_B(\varepsilon,T)$ is Bose distribution function. Evaluation of the integral (\ref{nn}) leads to the
result $T_c(\alpha) \approx 0.3 \Delta(\alpha)$ valid in the range $10 \leq \alpha < \alpha_c$. Note that
corrections to $T_c$ due to vortex depairing (Berezinsky-Kosterlitz-Thouless mechanism) are very weak,
due to smallness of the roton gap $\Delta$  in comparison with the plasma frequency $\omega_0$.

In conclusions, we have computed excitation spectrum of 2D Bose-liquid with long-range
interaction in a strong-coupling regime. Broad range of coupling strengths $\alpha < \alpha_c$
is found there gapful superfluid state is stable at $T=0$ in spite of a very small value of the roton gap $\Delta$.
Out data suggest a single quantum phase transition from such a strongly correlated superfluid into a quantum crystal
phase at $\alpha_c \approx 37 \div 38 $. At smaller $\alpha$, superfluid state is stable up to the critical temperature
$T_c \approx 0.3 \Delta(\alpha)$, which is orders of magnitude lower than a naive estimate $T_0 \sim \hbar^2 n/m$ would give.

We are grateful to L. B. Ioffe, D. A. Ivanov,  L. N. Shchur and M. A. Skvortsov for useful discussions and advises.
This research was supported by the RFBR grant \# 10-02-00554 and by the RAS Program  ``Quantum physics of condensed matter''.

\section{Supplementary online material}

\subsection{1. Mapping from quantum mechanics to classical stochastic evolution}

Standard Fokker-Planck equation corresponding to the Langevin dynamics, Eq.(\ref{Langevin}) is
\beq \dot P({\bf{r}}_1,..{\bf{r}}_N , t) = \sum\limits_{j}\frac{\partial }{\partial{\bf{r}}_j} (T\frac{\partial }{\partial{\bf{r}}_j} +
\frac{\partial V}{\partial{\bf{r}}_j})P({\bf{r}}_1,..{\bf{r}}_N , t)= \hat{W}P
\label{fplanck}\eeq
Equilibrium solution of Eq.(\ref{fplanck}) is given by $P_0=e^{-V/T}$.
 One can check that after the change of variables $P=\Psi_0({\bf{r}}_1,..{\bf{r}}_N)\Psi({\bf{r}}_1,..{\bf{r}}_N,t)$
  the equation (\ref{fplanck}) assumes the form of imaginary-time Shroedinger equation $\dot\Psi = \hat{H} \Psi$,
where Hamiltonian $H$ is constructed from the potential $V$ as shown in Eq.(\ref{H1}).
 For the following we denote a position in coordinate space $({\bf{r}}_1,.. {\bf{r}}_N) \equiv \varphi$
and will not use the specific form of $\hat{H}$.
The correspondence of classical and quantum correlation functions that we prove below is valid for any
  symmetric $H=H^T$ which ground state $\Psi_0(\varphi)$ is known exactly.
The symmetry condition $H =H^T$ leads to $H^*=H$, which enables us
to choose real wavefunctions, so  $P(\varphi,t)$ is always real.

Quantum states form a full system of orthogonal functions:
 \bea H \to \{ \Psi_\lambda(\varphi), \lambda\} \\ \hat{1} = \sum\limits_{\lambda}\Psi_\lambda(\varphi)\Psi_\lambda(\varphi') = \delta(\varphi -
 \varphi') \label{one}\eea
  Consider quantum correlation function:
 \beq C_q (t)= \langle \Psi_0 |  A e^{-iHt} B|\Psi_0 \rangle
\label{Cq}
 \eeq
 Inserting into R.H.S. of  Eq.(\ref{Cq}) the decomposition of the unity operator (\ref{one}) we obtain
\begin{eqnarray}
\label{cq}
C_q (t)=
\sum\limits_{\lambda}\langle \Psi_0 |  A|\lambda \rangle e^{-i\lambda t}\langle \lambda| B|\Psi_0 \rangle
= \\ \nonumber
\int d\varphi d\varphi' \sum\limits_{\lambda}\Psi_0(\varphi)\Psi_\lambda(\varphi)A(\varphi)
\Psi_0(\varphi')\Psi_\lambda(\varphi')B(\varphi') e^{-i\lambda t}
\end{eqnarray}
where $A$ and $B$ are diagonal operators (i.e. functions of coordinates $\varphi$ only).
The derivation of the quantum-clssical mapping begins with replacing variables
$P(\varphi,t) = \Psi_0(\varphi')\Psi(\varphi',t)$. The operator governing the classical stochastic
evolution is $W = -\Psi_0(\varphi) \hat H \frac{1}{\Psi_0(\varphi)} $. It's easy to see that $P_\lambda(\varphi) =
\Psi_0(\varphi')\Psi_\lambda(\varphi')$ are the  eigenfunctions for this operator, yet this system of eigenfunctions
is neither normalized nor orthogonal since the operator $W$ is a non-Hermetean one.
Combining the identity $H =H^T$ and the definition of $W$, we obtain: $P_0 W^T = W P_0$, which is the detailed balance
condition.

Rewriting Eq.(\ref{cq}) formally  in classical notations, we find
\beq C_q (t)=  \int d\varphi d\varphi' \sum\limits_{\lambda}P_\lambda(\varphi)A(\varphi) P_\lambda(\varphi')B(\varphi') e^{-i\lambda t}
\label{cqc}\eeq
Now we need to evaluate classical correlation function. We have the equation for probability density
$P(\varphi)$:
\beq \dot{P} =WP, \eeq
 and system is in the equilibrium state $P(t)=P_0$.
The two-time correlation function $C(\tau)$ (as given by R.H.S. of Eq.(\ref{Equiv})) is defined via
 stochastic process 2-time probability $P(\varphi,0,\varphi',\tau)$:
\beq C(\tau)= \int d\varphi d\varphi' P(\varphi,0,\varphi',\tau)A(\varphi) B(\varphi')
\label{Cq2}
\eeq
 where $P(\varphi,0,\varphi',\tau)=P(\varphi)p_{\varphi\to\varphi'}^{\tau}$ according to the definition of
a conditional probability $p_{\varphi\to\varphi'}^{\tau}$ that the system will be in configuration $\varphi'$ at time $\tau$, given that it was
in configuration $\varphi$ at time $\tau=0$. Substituting this expression for $P(\varphi,0,\varphi',\tau)$ into Eq.(\ref{Cq2})
we find:
 \beq
C(\tau) =\int d\varphi d\varphi' P_0(\varphi) A(\varphi)p_{\varphi\to\varphi'}^{\tau} B(\varphi')\eeq
 To evaluate
$p_{\varphi\to\varphi'}^{\tau}=(e^{W\tau})_{\varphi'}\delta(\varphi' -\varphi)$, we need to know the decomposition of $\delta$- function into
eigenmodes. it is convinient to use "quantum" basis (since classical operator $W$ is non-Hermetean):
\beq
\delta(\varphi' -\varphi) =
\sum\limits_{\lambda}\Psi_\lambda(\varphi)\Psi_\lambda(\varphi') =
\frac{\sum\limits_{\lambda}P_\lambda(\varphi)P_\lambda(\varphi')}{P_0(\varphi)}
\label{Cq3}
\eeq
%
Now we can contract this $\delta$-function with $e^{W\tau}$ (remember that eigenvalues are $-\lambda$)
\beq p_{\varphi\to\varphi'}^{\tau}=(e^{W\tau})_{\varphi'}\delta(\varphi' -\varphi) =
\frac{\sum\limits_{\lambda}P_\lambda(\varphi)P_\lambda(\varphi')e^{-\lambda \tau}}{P_0(\varphi)}\eeq
 For classical correlation function we
obtain
\bea
\label{Cq4}
 C(\tau)= \int d\varphi d\varphi' P_0(\varphi) A(\varphi)\frac{\sum\limits_{\lambda}P_\lambda(\varphi)P_\lambda(\varphi')e^{-\lambda\tau}}{P_0(\varphi)} B(\varphi') = \\ \nonumber
= \sum\limits_{\lambda}P_\lambda(\varphi)A(\varphi)P_\lambda(\varphi')B(\varphi')e^{-\lambda \tau}
\eea
Comparing Eqs.(\ref{Cq4}) and (\ref{cqc}) we find the relation wanted:
 \beq C_q (t)=C(it)
\eeq
%

\subsection{2. Details of data analysis.}

 For rotonic spectrum with gap $\Delta$ quasiparticle continuum begins at $\omega>2\Delta$. It can be seen by considering 2
rotons with minimal energy ($k_{1,2}=k_0$) and arbitrary angle between $k_1$ and $k_2$. Total energy is $2\Delta$, and total momentum can be set
arbitrary in the region $K<2k_0$. Rotons are the only detected excitations below the continuum:
\bea S(\omega, k) = A \delta(\omega - E(k)) + S_{con}(\omega ,k), \label{prr}\\ \textrm{where }~ ~ ~ S_{con}(\omega<2\Delta)=0, \\ E(k)\approx
\Delta + \frac{(k-k_0)^2}{2m^*} ~ ~ \textrm{at} ~ ~{k\to k_0}\eea
 In the region $k\in [p_-, p_+]$ we assume the main contribution to come from a quasiparticle, i.e. in Eq.(\ref{prr})
 \beq A\gg \int S_{con}(\omega,k) d\omega, \label{amp}\eeq
 so that the exact shape of $S_{con}(\omega)$ does not matter. For data fitting we use rectangular spectral density $S_{con}(\omega)= B, ~ ~ 2\Delta<
 \omega<\omega_{max}$, so for each value of $k$  there are four fiting  parameters: $A, E(k), B, \omega_{max}$,
apart from the value of $\Delta=E(k_0)$ that is the same for all $k$.
We minimize the mean  square deviation $\sum_i(S_{fit}(t_i)-S_{sim}(t_i))^2$ to find $E(k)$ plotted on Fig~\ref{Fig:a20}. We also check the condition (\ref{amp})
 and find that it is violated in the close vicinity of terminating points, thus the statistical error of determining $E(k)$ grows there.

To collect data presented in  Fig~\ref{Fig:gap1}, we do not need to use the $k$-regions near the terminating points $k\approx p_-, p_+$ , so we
can  use inequality (\ref{amp}) and estimate $E(k)$ just as $d\hbox{ln}\mathcal{S}/dt |_{t=t_o}$ where $t_o$ is sufficiently long to lead to
additional exponential damping of the continuum modes. Note that inaccuracy in determination of $\hbox{ln}\mathcal{S}$ (and  of its derivative)
grows exponentially with $t_o$, since $1/\mathcal{S}\sim e^{E(k)t_o}$. Therefore the finite simulation time determines  how long is the optimal
interval $t_o$ we can use. The derivative $ d\hbox{ln}\mathcal{S}/d t$ can be accessed with the use of Monte Carlo estimator (subtracting the
$\mathcal{S}$ values for consequent configurations), or by drawing a line through the sequence of points $\hbox{ln}\mathcal{S}(t_i),~ ~
t_i\in[t_o- \delta T/2,t_o+ \delta T/2 ]$ . These approaches yield similar results, but the latter is more insightful when one tries to assign
errorbars $\sigma E(k)$ to the  results for $E(k)$. These errors contain standard N-point slope measurement error $\sigma_N E =
\sqrt{\frac{12}{N}}\frac{\sigma\hbox{ln}\mathcal{S}}{\delta T}$ and the systematical overestimating of $E(k)$ due to the continuum modes. The
second source if errors is related with the fact  that $\hbox{ln}\mathcal{S}(t_i)$ is not exactly linear function of time. Assuming that the
derivative $ d\hbox{ln}\mathcal{S}/d t$ changes by $\delta E$ in the interval $\delta T$, we can estimate  possible systematic errors as
$\sigma_c E= \delta E e^{-(2\Delta -E)\delta T/2}/(1- e^{-(2\Delta -E)\delta T})$. The denominator of this expression diverges while $k$
approach terminating points, which reminds us of the range of applicability of the method we used.
 Surprisingly, the data analysis using N-point treatment of $\hbox{ln}\mathcal{S}(t)$ and
neglecting  systematic shift $\sigma_c E$, can be performed in the whole range of $k \in (p_-, p_+)$.
This method catches non-analitic behaviour of $E(k)$ near the  spectrum terminating points, as well
vanishing of the roton gap $\Delta$ while $\alpha$ approaches $\alpha_c$.
In both these cases,  errors bars $\sigma_N E$ grow considerably, indicating the approach to a transition.

\end{document}